\documentstyle[12pt]{article}

\begin{document}
\title{Covariance of noncommutative Gra\ss mann star product}
\author{{\bf M. Daoud} \\
Abdus Salam International Centre of Theretical Physics \\
Trieste, Italy\\
and
\\
University Ibn Zohr, LPMC , Department of physics \\  Agadir, Morocco}
\maketitle

\begin{abstract}
 Using the Coherent states of many fermionic degrees of freedom labeled by Gra\ss mann variables, we introduce the noncommutative (precisely non anticommutative) Gra\ss mann star product. The covariance of star product under unitary transformations, particularly canonical ones, is studied. The super star product, based on supercoherent states of supersymmetric harmonic oscillator, is also considered.
\end{abstract}

\vfill
\newpage 

\section{Introduction}
\hspace*{0.5cm} In 1949, Moyal introduced a new bracket for functions on the classical phase space that replaces the Poisson one in the quantization procedure [1]. This bracket is closely related to Weyl's correspondence rule between classical and quantum observables [2]. Recently, deformation \`a la Moyal has been explored in several context as noncommutative approach to string theory [3], matrix model [4], the non-commutative Yang-Mills theories [5] and non-commutative gauge theories [6]. More generally, models on noncommutative space time, obtained by replacing the ordinary product of functions by the noncommutative Moyal product, helped to revive the deformation quantization technique which was elucited further in [7]. Throught this article, we will deal with the concept of coherent states which as it is well established have important application building a bridge between the classical and the quantum worlds views [8,9]. In other side, coherent states have fundamental properties like over-completion and non-orthogonality which constitute the main ingredients to formulate the Moyal product in coherent state framework. The first result obtained in this sens is due to Alexanian and al [10](see also [11]). Coherent states  of quantum systems with nonlinear spectrum has been also considered  to find new kinds of star products [12] and ones related to $su(2)$ algebra has been nicely used to introduce a new star product on the fuzzy sphere [13]. \\
\hspace*{0.5cm}The main task of this work is to derive star product for gra\ss mannian manifold using the fermionic coherent states. Section 2 collects some notes on Gra\ss mann variables [14] and coherent states of a system with finite number of fermions [15]. Section 3 develops a consistant star product using coherent states associated to fermionic degrees of freedom. Covariance with respect the unitary transformation is studied in section 4. It will be shown that the covariance is not guaranteed under arbitrary unitary transformation except in the particular case of canonical ones. In section 5, the correspondence rules between functions and operators are used to generate the "star"-analogue of the supersymmetric oscillator. Concluding remarks close this note.
\section{\bf Basic concepts of fermionic coherent states}
\subsection{\bf Gra\ss mann variables}
\hspace*{0.5cm}It is well established that the Gra\ss mann variables serve as classical analogues of fermionic degrees of freedom. Indeed, the phase space, in the classical sens, of a collection of $N$ fermions can be identified with the Gra\ss mann algebra ${\bf C}B_{2N}$ which is generated by a set of $2N$ independants Gra\ss mann variables obeying the anti-commutation relation
\begin{equation}
 \{ \theta_{i} , \overline{\theta_{j}} \} = \delta_{ij}{\hskip 1cm} \{ \theta_{i} , \theta_{j} \} =\{ \overline{ \theta_{i}} , \overline{\theta_{j}} \} = 0
\end{equation}
where $i = 1, 2,...,N$. This algebra is ${\bf Z}_2$-graded; i.e the even elements of ${\bf C}B_{2N}$ are commuting and have even Gra\ss mann parity. The odd elements are anti-commuting with odd parity. The Gra\ss mann integration and differentiation are given by
\begin{equation}
\int d\theta = 0 {\hskip 1cm} \int d\theta \theta= 0 {\hskip 1cm} \frac{d}{d\theta} 1 = 0 {\hskip 1cm} \frac{d}{d\theta} \theta = 1
\end{equation}
where $\theta$ stand for $\theta_i$ and $\overline\theta_i$ $( i= 1, 2,...,N )$. Note that the integration and differentiation are treated like odd Gra\ss mann quantities according to ${\bf Z}_2$-graduation (for more details see [14]).
\subsection{\bf Fermionic coherent states}
\hspace*{0.5cm} We consider a finite number $N$ of fermionic degrees of freedom which are characterized by annihilation and creation operators $f_{i}^-$ and $f_{i}^+ = (f_{i}^-)^+$ ($ i = 1, 2, ...,N$) satisfying the canonical anti-commutation relations
\begin{equation}
\{f_{i}^- , f_{j}^- \} = \{f_{i}^+ , f_{j}^+ \} = 0 {\hskip 1cm} \{f_{i}^- , f_{j}^+ \}= \delta_{ij}
\end{equation}
The corressponding Hilbert space is the $N$-fold tensor product $ {\cal H}_f = {\cal H}_{1} \otimes {\cal H}_{2} \otimes .... \otimes{\cal H}_{N}$ of two dimensional Fock space
\begin{equation}
{\cal H}_{i} = \bigg\{ \vert 0 \rangle_{i} ,  \vert 1 \rangle_{i} \bigg\}
\end{equation}
The operators $f_{i}^-$ act on ${\cal H}_{i}$ as
\begin{equation}
f_{i}^{-} \vert 0 \rangle_{i} = 0 {\hskip 1cm} f_{i}^{-} \vert 1
\rangle_{i} =  \vert 0 \rangle_{i}
\end{equation}
The Hilbert space ${\cal H}_{f}$ is spanned by the eigenstates
\begin{equation}
\vert n_{1},n_{2},...,n_{N}\rangle \equiv \vert n_{1}\rangle_1\otimes \vert n_{2}\rangle_2\otimes ...\otimes\vert n_{N}\rangle_N
\end{equation}
The fermionic coherent states are defined in analogy with the standard canonical harmonic oscillator. They are defined as eigenstates of annihilation operators $f_{i}^-$ and are labeled by Gra\ss mann variables. They can be  defined also as the action of a displacement operator on the ground state of the fermionic system under consideration. The coherent state  of one fermionic degree of freedom is given by
\begin{equation}
\vert \theta \rangle \equiv \exp ( \theta f^{+} + \bar {\theta} f^-) \vert 0 \rangle
\end{equation}
Using the action of  annihilation and creation operators on the one fermion Hilbert space, one has
\begin{equation}
\vert \theta \rangle = ( 1 - \frac{1}{2}\bar {\theta} \theta)[ \vert 0 \rangle + \theta \vert 1 \rangle] = \exp (\frac{1}{2}\bar {\theta} \theta)[ \vert 0 \rangle + \theta \vert 1 \rangle]
\end{equation}
Note that $\vert \theta \rangle$ is eigenstate of the operator $ f^-$ with eigenvalue $\theta$. The normalized states (8) form an overcomplete set and provide a resolution of the identity
\begin{equation}
\int d\mu( \theta, \bar {\theta}) \vert \theta \rangle \langle \theta \vert = 1
\end{equation}
where the measure is defined by $ d\mu( \theta, \bar {\theta}) = d\bar {\theta}d\theta$.\\
For $N$ fermionic degrees of freedom, we define the coherent states, as tensorial product of the state $ \vert \theta_i \rangle_{i} $ of the fermion $i$
\begin{equation}
\vert \vec {\theta}\rangle = \vert \theta_{1} \rangle_{1} \otimes \vert \theta_{2} \rangle_{2} \otimes \vert \theta_{N} \rangle_{N}
\end{equation}
where
\begin{equation}
\vert \theta_{i} \rangle_{i} = ( 1 - \frac{1}{2}\bar {\theta_{i}} \theta_{i})[ \vert 0 \rangle_{i} + \theta_{i} \vert 1 \rangle_{i}]
\end{equation}
The coherent states $\vert \vec {\theta}\rangle$ form an overcomplete basis for the Hilbert space ${\cal H}_f$ and the completeness relation is given by
\begin{equation}
\int d\mu(\vec {\theta}, \vec {\bar {\theta}}) \vert \vec {\theta} \rangle \langle \vec {\theta} \vert = 1
\end{equation}
where the measure is given now by
\begin{equation}
d\mu(\vec {\theta}, \vec {\bar {\theta}}) = \prod_{i=1}^{N}  d\bar {\theta}_{i}d\theta_{i}.
\end{equation}
\section{\bf Gra\ss mann star product }
\hspace*{0.5cm} The main ingredient to introduce the Gra\ss mann star product is the set of the fermionic coherent states $\vert \vec{\theta} \rangle$. For this end, let us consider an operator $A$, acting on the Hilbert space ${\cal H}_{f}$, which can be expanded in terms of ladder fermionic operators as
\begin{equation}
A = \sum_{\vec {m} , \vec {n}}
a_{\vec {m} , \vec {n}} f_{1}^{+n_{1}}f_{2}^{+n_{2}}...f_{N}^{+n_{N}} f_{1}^{-m_{1}}f_{2}^{-m_{2}}...f_{N}^{-m_{N}}
\end{equation}
where $\vec {m} = (m_{1},m_{2},...,m_{N})$ and $\vec {n} = (n_{1},n_{2},...,n_{N})$ are $N$-dimensional vectors.  According the nilpotency of the fermionic operators, the  admissible values of integers $m_{i}$ and $n_{i}$ are 0 or 1. In the fermionic coherent states representation, we associate to the operator $A$ the following function
\begin{equation}
{\cal A}(\vec {\theta}, \vec {\bar \theta}) = \langle \vec {\theta} \vert A  \vert \vec {\theta} \rangle
= \sum_{\vec {m} , \vec {n}}
a_{\vec {m} , \vec {n}}
 \bar {\theta}_{1}^{n_{1}}\bar {\theta}_{2}^{n_{2}}...\bar {\theta}_{N}^{n_{N}}
 {\theta}_{1}^{m_{1}} {\theta}_{2}^{m_{2}}... {\theta}_{N}^{m_{N}}
\end{equation}
in terms of the Gra\ss mann variables $\theta_{i}$ and $\bar \theta_{i}$$(i = 1, 2,...,N)$. As in the bosonic case [10-12], an associative product of two functions is then defined by
\begin{equation}
{\cal A}(\vec {\theta}, \vec {\bar \theta}) \star {\cal B}(\vec {\theta}, \vec {\bar \theta}) = \langle \vec {\theta} \vert AB \vert \vec {\theta} \rangle
\end{equation}
The associativity of this star product originates from the associativity of the algebra generated by fermionic creation and annihilation operators.
Using the overcompletion property (12), the star product (16) can be written as
\begin{equation}
{\cal A}(\vec {\theta}, \vec {\bar \theta}) \star {\cal B}(\vec {\theta}, \vec {\bar \theta}) =
\int d\mu(\vec {\eta}, \vec {\bar {\eta}}) \langle \vec {\theta} \vert A \vert  \vec {\eta}  \rangle \langle \vec {\eta} \vert B \vert \vec {\theta} \rangle
\end{equation}
Remark that the matrix elements $\langle \vec {\theta} \vert A \vert \vec {\eta} \rangle$ can be expressed as the action of a translation operator on the function ${\cal A}(\vec {\theta}, \vec {\bar \theta})$
\begin{equation}
e^{-\theta_{i}\frac{\partial}{\partial\eta_{i}}}e^{\eta_{i}\frac{\partial}{\partial\theta_{i}}}{\cal A}(\vec {\theta}, \vec {\bar \theta}) = \frac{\langle \vec {\theta} \vert A \vert \vec {\eta} \rangle}{\langle \vec {\theta} \vert  \vec {\eta} \rangle}
\end{equation}
and
\begin{equation}
e^{-\theta_{i}\frac{\partial}{\partial\bar {\eta_{i}}}}e^{\bar {\eta_{i}}\frac{\partial}{\partial\theta_{i}}}{\cal A}(\vec {\theta}, \vec {\bar \theta}) = \frac{\langle \vec {\eta} \vert A \vert \vec {\theta} \rangle}{\langle \vec {\eta} \vert  \vec {\theta} \rangle}
\end{equation}
where we have summation over repetead indices. The translation operators appearing in (18) and (19) can be written as ordered exponential
\begin{equation}
: e^{(\eta_i - \theta_i)\overrightarrow{\frac{\partial}{\partial\theta_i}}}:
\end{equation}
and
\begin{equation}
: e^{\overleftarrow{\frac{\partial}{\partial\theta_i}}(\eta_i - \theta_i)}:
\end{equation}
respectively. So, the Gra\ss man star product (16) is given now by
\begin{equation}
\star = \int d\mu(\vec {\eta} ,\vec {\bar \eta})
: e^{\overleftarrow{\frac{\partial}{\partial\theta_i}}(\eta_i - \theta_i)}:
\vert \langle \vec {\theta} \vert  \vec {\eta} \rangle \vert^2
: e^{(\bar {\eta_i} - \bar {\theta_i})\overrightarrow{\frac{\partial}{\partial \bar {\theta_i}}}}:
\end{equation}
From the equations (10) and (11), the overlapping of two fermionic coherent states is
\begin{equation}
\vert \langle \vec {\theta} \vert  \vec {\eta} \rangle \vert^2 = e^{-(\overline{\eta_i - \theta_i})(\eta_i - \theta_i)}
\end{equation}
Substituting (23) in (22) and computing the integral (22) with a simple change of variables, we obtain
\begin{equation}
\star = e^{\overleftarrow{\frac{\partial}{\partial\theta_i}} \overrightarrow{\frac{\partial}{\partial \bar \theta_i}}}
\end{equation}
This is the Gra\ss mann star product. The computation of the star product of two arbitrary functions ${\cal A}(\vec {\theta}, \vec {\bar \theta}) $ and ${\cal B}(\vec {\theta}, \vec {\bar \theta})$ becomes easy thanks to correspondence between operators acting on the fermionic Hilbert space and functions of Gra\ss mannian variables (Eq.(15)) and the minimal set of
the needed relations, to perform the star calculus, are
\begin{equation}
\bar {\theta_{i}} \star \theta_{j} = \bar {\theta_{i}} \theta_{j}
\end{equation}
\begin{equation}
\theta_{i} \star \theta_{j} =  \theta_{i} \theta_{j}
\end{equation}
\begin{equation}
\bar {\theta_{i}} \star \bar {\theta_{j}} = \bar {\theta_{i}} \bar {\theta_{j}}
\end{equation}
\begin{equation}
\theta_{i} \star \bar {\theta_{j}} = \theta_{i} \bar {\theta_{j}} + \delta_{ij}
\end{equation}
Notice that in the star formulation, we have the following star anti-commutation
\begin{equation}
\{ \theta_{i} , \theta_{j} \}_{\star} = \{ \bar {\theta_{i}} , \bar {\theta_{j}}\}_{\star} = 0 $$\vskip 0.005cm$$
\{ \bar {\theta_{i}} , \theta_{j}\}_{\star} = \delta_{ij}
\end{equation}
which are the star analogue of the anti-commutation relation of the fermionic operators. Note also that the nilpotency of Gra\ss mann variables
\begin{equation}
\theta_{i} \star \theta_{i} = 0
\end{equation}
is preserved in the star language.
\section{Covariance of Gra\ss mann star product under unitary transformation}
\hspace*{0.5cm} In this section, we analyse the covariance of Gra\ss mann star product under unitary transformation. In this order, we consider  the unitary transformation of the operator $A$ under $ O = e^{\Lambda}$ such that $\Lambda^{+} + \Lambda = 0$
\begin{equation}
A' = O^{+}  A  O = e^{-\Lambda}  A  e^{\Lambda}$${\vskip 0.05cm}$$
= \sum_{k=0}^{\infty} \frac{(-1)^{k}}{k!}\bigg[ \Lambda, \big[\Lambda ,...,[\Lambda , A ]\big] \bigg]
\end{equation}
The function associated to the operator $A'$ is given by
\begin{equation}
{\cal A'}(\vec {\theta}, \vec {\bar \theta}) =  \langle \vec {\theta} \vert A'  \vert \vec {\theta} \rangle $${\vskip 0.05cm}$$
= \sum_{k=0}^{\infty} \frac{(-1)^{k}}{k!}\bigg\{ \lambda, \big\{\lambda ,...,\{\lambda , A \}_{\star}\big\}_{\star} \bigg\}_{\star}
\end{equation}
where $\lambda$ is the function associated to the anti-hermitian operator $\Lambda$
\begin{equation}
\lambda \equiv \lambda(\vec {\theta}, \vec {\bar \theta}) =  \langle \vec {\theta} \vert {\lambda}  \vert \vec {\theta} \rangle
\end{equation}
In the noncommutative Gra\ss mann space, the function ${\cal A'}$ can be written also as
\begin{equation}
{\cal A'}(\vec {\theta}, \vec {\bar \theta}) = e^{-{\cal D}_{\lambda}}{\cal A}(\vec {\theta}, \vec {\bar \theta})
\end{equation}
where the action of ${\cal D}_{\lambda}$ on the functions defined on the noncommutative Gra\ss mann space is defined by
\begin{equation}
{\cal D}_{\lambda}{\cal A}(\vec {\theta}, \vec {\bar \theta})
 = \lambda(\vec {\theta}, \vec {\bar \theta}) \star {\cal A}(\vec {\theta}, \vec {\bar \theta})-
{\cal A}(\vec {\theta}, \vec {\bar \theta}) \star  \lambda(\vec {\theta}, \vec {\bar \theta})
\end{equation}
Hence, the unitary transformation $ O = e^{\Lambda}$ in the operator space has unique representation in the noncommutative Gra\ss mann space formulated by equation (34). One can verify also that
\begin{equation}
e_{\star}^{-\lambda} \star e_{\star}^{\lambda} = e_{\star}^{\lambda} \star e_{\star}^{-\lambda} = 1
\end{equation}
where the star exponontial is defined as follows
\begin{equation}
e_{\star}^{x} = \sum_{k=0}^{\infty}\frac{1}{k!} (x^{\star})^k
\end{equation}
Remark that for two transformations $ O_{1} = e^{\Lambda_{1}}$ and $O_{2} = e^{\Lambda_{2}}$ acting on the operator space, one has
\begin{equation}
[{\cal D}_{\lambda_1} , {\cal D}_{\lambda_2}] = {\cal D}_{\{\lambda_{1},\lambda_{2}\}_{\star}}
\end{equation}
where the functions $\lambda_1$ and $\lambda_2$ are ones associated to anti-hermitians operators
$\Lambda_1$ and $\Lambda_2$, respectively.\\
Using the previous resuls, one can write the functions ${\cal A'}$ in the form
\begin{equation}
{\cal A'}(\vec {\theta}, \vec {\bar \theta}) = \sum_{\vec {m} , \vec {n}}
a_{\vec {m} , \vec {n}}
\bar {\psi}_{1}^{n_{1}}\star\bar {\psi}_{2}^{n_{2}}\star...\star\bar {\psi}_{N}^{n_{N}}\star
{\psi}_{1}^{m_{1}} \star{\psi}_{2}^{m_{2}}\star... \star{\psi}_{N}^{m_{N}}
\end{equation}
where the new variables $\psi$ are related to old ones as follows
\begin{equation}
\psi_i = e^{-{\cal D}_\lambda}\theta_{i} {\hskip 1cm} \bar \psi_i = e^{-{\cal D}_\lambda}\bar \theta_{i}
\end{equation}
It is clear that, for an arbitrary unitary transformation, we have
\begin{equation}
{\cal A'}(\vec {\theta}, \vec {\bar \theta}) \ne {\cal A}(\vec {\psi}, \vec {\bar \psi})
\end{equation}
and the covariance is violated. A necessary condition to get a covariance is
\begin{equation}
e^{\overleftarrow{\frac{\partial}{\partial\psi_i}} \overrightarrow{\frac{\partial}{\partial \bar \psi_i}}} = e^{\overleftarrow{\frac{\partial}{\partial\theta_i}} \overrightarrow{\frac{\partial}{\partial \bar \theta_i}}}
\end{equation}
with summation over the repeated indices. This situation occurs in the exceptional case in which the transformation $O$ belongs to the group of real orthogonal $2N\times2N$ matrices, $SO(2N,R)$. To clarify how the classical covariance is preserved under canonical transformations belonging to $SO(2N,R)$, let us consider linear canonical transformations;i.e, the linear homogenous transformations of the fermionic operators $f_{i}^-$ and $f_{i}^+$, that do not change the commutations realtions (3)
\begin{equation}
F_{i}^{-} = u_{ij} f_{j}^{-} + v_{ij} f_{j}^{+}
\end{equation}
which can written in compcat matricial form as follows
\begin{equation}
F^{-} = U f^{-} + V f^{+}
\end{equation}
The matrix elements of $U$ and $V$ are $u_{ij}$ and $v_{ij}$ respectively.
The necessary relations for the transformation (44) to be canonical are
\begin{equation}
U V' + V U' = 1 {\hskip 1cm} U U^{+} + V V^{+} = 1
\end{equation}
where ' stands for matrix tranposition and $^+$ for hemitian conjugason.\\
According to Stone-Von Neuman theorem (see [8] for instance), there exist an unitary operator $O$ such that the operators $F_{i}^{-}$ and $f_{i}^{-}$ are equivalents
\begin{equation}
F_{i}^{-} = O^{+} f_{i}^{-} O
\end{equation}
It is given by
\begin{equation}
O = e^{( \frac {\alpha_{ij}}{2}f_{i}^{+}f_{j}^{+} + \frac {\overline {\alpha_{ij}}}{2}f_{i}^{-}f_{j}^{-})} = e^\Lambda
\end{equation}
where $\alpha_{ij}$ are elements of complex anti-symmetrical matrix $\alpha$. A direct computation
leads to
\begin{equation}
F^{-} = \cos\sqrt{\alpha^{+}\alpha}f^{-} + \frac{\alpha}{\sqrt{\alpha^{+}\alpha}}\sin\sqrt{\alpha^{+}\alpha}f^{+}
\end{equation}
\begin{equation}
F^{+} = \cos\sqrt{\alpha^{+}\alpha}f^{+} + \frac{\alpha^+}{\sqrt{\alpha^{+}\alpha}}\sin\sqrt{\alpha^{+}\alpha}f^{-}
\end{equation}
Using the operator-function correspondence rule (15), the function associated to the operator $\Lambda$ is
\begin{equation}
\lambda \equiv  \lambda (\vec {\theta},\vec {\bar \theta})= \frac {\alpha_{ij}}{2}\bar {\theta}_{i}\bar {\theta}_{j} + \frac {\overline {\alpha_{ij}}}{2}\theta_{i}\theta_{j}
\end{equation}
The action of the operator ${\cal D}_\lambda$ on the canonical Gra\ss mann variables gives
\begin{equation}
{\cal D}_{\lambda} \theta_{k} = - \alpha_{ki} \bar {\theta}_{i} {\hskip 1cm}
{\cal D}_{\lambda} \bar {\theta_{k}} =  \overline {\alpha_{jk}} \theta_{j}
\end{equation}
which can be written in a compact form as
\begin{equation}
{\cal D}_{\lambda} \theta = - \alpha  \bar {\theta} {\hskip 1cm} {\cal D}_{\lambda} \bar {\theta}= \alpha^{+}\theta,
\end{equation}
 and from which one get
\begin{equation}
e^{-{\cal D}_{\lambda}}\theta = \cos\sqrt{\alpha^{+}\alpha} \theta + \frac{\alpha}{\sqrt{\alpha^{+}\alpha}}\sin\sqrt{\alpha^{+}\alpha} \bar {\theta} = \psi
\end{equation}
\begin{equation}
e^{-{\cal D}_{\lambda}} \bar\theta = \cos\sqrt{\alpha^{+}\alpha}\bar \theta + \frac{\alpha}{\sqrt{\alpha^{+}\alpha}}\sin\sqrt{\alpha^{+}\alpha} \theta = \bar \psi
\end{equation}
At this stage, it is interesting to note that it is allowed to label coherent states associated to the new fermionic algebra $\{F_{i}^{-}, F_{i}^{+}\}$ by the variables $\psi_i$ and $\bar \psi_i$. Then, by virtue of the correspondence rule (15), the variables $\psi_i$ (resp. $\bar \psi_i$) are associated to annihilation operators $F_{i}^{-}$ (resp. creation operators $ F_{i}^{+}$) and  following the analysis, presented in section 2, a new star product can be defined as
\begin{equation}
\tilde \star =
e^{\overleftarrow{\frac{\partial}{\partial\psi_i}} \overrightarrow{\frac{\partial}{\partial \bar \psi_i}}},
\end{equation}
and we have
\begin{equation}
\bar {\psi_{i}} \tilde\star \psi_{j} = \bar {\psi_{i}} \psi_{j}
\end{equation}
\begin{equation}
\psi_{i} \tilde\star \psi_{j} =  \psi_{i} \psi_{j}
\end{equation}
\begin{equation}
\bar {\psi_{i}} \tilde\star \bar {\psi_{j}} = \bar {\psi_{i}} \bar {\psi_{j}}
\end{equation}
\begin{equation}
\psi_{i} \tilde\star \bar {\psi_{j}} = \psi_{i} \bar {\psi_{j}} + \delta_{ij}
\end{equation}
The new canonical variables $\psi_{i}$ and $\bar {\psi_{j}}$ satisfy the following star anti-commutation
\begin{equation}
\{ \psi_{i} , \psi_{j} \}_{\tilde\star} = \{ \bar {\psi_{i}} , \bar {\psi_{j}}\}_{\tilde\star} = 0 $$\vskip 0.005cm$$
\{ \bar {\psi_{i}} , \psi_{j}\}_{\tilde\star} = \delta_{ij}
\end{equation}
Finally, one conclude that, under canonical transformations, the anti-commutations relations of Gra\ss mann algebra are preserved, the products  $\star$ and $\tilde\star$ are identical and the covariance is guaranteed
\begin{equation}
{\cal A'}(\vec {\theta}, \vec {\bar \theta}) = e^{-{\cal D}_{\lambda}}{\cal A}(\vec {\theta}, \vec {\bar \theta}) = {\cal A}(\vec {\psi}, \vec {\bar \psi})
\end{equation}
contrary to the case of arbitrary unitary transformations.
\section{\bf Super $\star$-analogue of the supersymmetric oscillator}
\hspace*{0.5cm} We start by examining the simplest case of fermionic oscillator. The Hamiltonian is given by
\begin{equation}
H_f = f^{+}f^{-}
\end{equation}
where $f^{+}$ and $f^{-}$ are creation and annihilation operators ( the eigenstates are $\vert 0 \rangle$ and $\vert 1 \rangle$ with eigenvalues $e_{0} = 0$ and $e_{1} = 1$, respectively). Using the fermionic coherent states (8), the function associated to the Hamiltonian $H$ is given by
\begin{equation}
h(\theta,\bar {\theta}) = \langle \theta \vert H \vert \theta \rangle = \bar {\theta}\theta
\end{equation}
Any operator, acting on the fermionic Hilbert space, can be expanded in term of the elements of the occupation number operators basis $\{P_{m,n} = \vert m \rangle \langle n \vert\}$. The functions associated to elements of this basis are defined by
\begin{equation}
{\cal P}_{m,n}(\theta,\bar {\theta}) = \langle \theta \vert m \rangle \langle n \vert \theta \rangle
\end{equation}
The function ${\cal P}_{0,0}(\theta,\bar {\theta})$ provide a $\star$-vacuum. Indeed, we have
\begin{equation}
\theta \star {\cal P}_{0,0}(\theta,\bar {\theta}) = {\cal P}_{0,0}(\theta,\bar {\theta}) \star \bar\theta = 0
\end{equation}
Furthermore, we have the following completion relation
\begin{equation}
{\cal P}_{0,0}(\theta,\bar {\theta}) + {\cal P}_{1,1}(\theta,\bar {\theta})= 1
\end{equation}
which is the analogue of the completion relation $\vert 0 \rangle \langle 0 \vert + \vert 1 \rangle \langle 1 \vert = 1$. One can verify also that
\begin{equation}
h(\theta,\bar {\theta})\star {\cal P}_{m,n}(\theta,\bar {\theta}) = e_{m} {\cal P}_{m,n}(\theta,\bar {\theta})
\end{equation}
\begin{equation}
{\cal P}_{m,n}(\theta,\bar {\theta}) \star h(\theta,\bar {\theta})  = e_{n} {\cal P}_{m,n}(\theta,\bar {\theta})
\end{equation}
\begin{equation}
\theta \star {\cal P}_{m,n}(\theta,\bar {\theta}) = {\cal P}_{m-1,n}(\theta,\bar {\theta})( 1- \delta_{m,0})
\end{equation}
\begin{equation}
 {\cal P}_{m,n}(\theta,\bar {\theta})\star \theta  = {\cal P}_{m,n+1}(\theta,\bar {\theta})( 1- \delta_{n,1})
\end{equation}
\begin{equation}
\bar {\theta} \star {\cal P}_{m,n}(\theta,\bar {\theta}) = {\cal P}_{m+1,n}(\theta,\bar {\theta})( 1- \delta_{m,1})
\end{equation}
\begin{equation}
 {\cal P}_{m,n}(\theta,\bar {\theta})\star \bar {\theta}  = {\cal P}_{m,n-1}(\theta,\bar {\theta})( 1- \delta_{n,0})
\end{equation}
It is interesting to note that the functions ${\cal P}_{m,n}(\theta,\bar {\theta})$ are related to the generating function for the fermionic oscillator:
\begin{equation}
\Phi(\alpha, \beta,\theta,\bar \theta) = \sum_{mn}\alpha^{n}\beta^{m} \langle n \vert \exp(\theta f ^{+} + \bar {\theta} f^{-})\vert m \rangle
\end{equation}
where the integers $m$ and $n$ take the values 0, 1 and $\alpha$ and $\beta$ are Gra\ss mann variables. Computing the matrix elements occuring in the last equation, we have
\begin{equation}
\Phi(\alpha, \beta,\theta,\bar \theta) = \exp ( -\frac{1}{2}(\bar {\theta}\theta + \alpha\theta + \beta\bar {\theta} + \alpha\beta))
\end{equation}
One can verify
\begin{equation}
\frac{\partial^n}{\partial\alpha^n}\frac{\partial^m}{\partial\alpha^m}\Phi(\alpha, \beta,\theta,\bar \theta)\vert_{\alpha=\beta=0} = {\cal P}_{m,n}(\theta,\bar {\theta})\star \bar {\theta}
\end{equation}
traducing the relation between the generating function and the functions associated to elements $P_{m,n}$.\\
The Gra\ss mann star product, defined here by means of fermionic coherent states, is equivalent to the following one
\begin{equation}
\hat \star = e^{ \frac{1}{2}\big(\overleftarrow{\frac{\partial}{\partial\theta}} \overrightarrow{\frac{\partial}{\partial \bar \theta}} + \overleftarrow{\frac{\partial}{\partial\bar \theta}} \overrightarrow{\frac{\partial}{\partial  \theta}}\big)}
\end{equation}
which seems to be  the Gra\ss mann version of Moyal product in the noncommutative Gra\ss mann space. The equivalence relation is
\begin{equation}
T({\cal A}(\vec {\theta}, \vec {\bar \theta}))\star T({\cal B}(\vec {\theta}, \vec {\bar \theta})) = T({\cal A}(\vec {\theta}, \vec {\bar \theta})\hat \star {\cal B}(\vec {\theta}, \vec {\bar \theta}))
\end{equation}
where the relevant operator is
\begin{equation}
T = e^{\frac{1}{2}\frac{\partial}{\partial\theta}\frac{\partial}{\partial\bar \theta}}= 1 - \frac{1}{2}\frac{\partial}{\partial\theta}\frac{\partial}{\partial\bar \theta}
\end{equation}
The needed relations to compute the $\hat \star$ product of two arbitrary functions are
\begin{equation}
\theta \hat \star \bar {\theta} = \theta \bar {\theta} + \frac{1}{2} {\hskip 2cm}
\bar {\theta} \hat \star \theta = \bar {\theta}\theta  + \frac{1}{2}
\end{equation}
from which one has
\begin{equation}
\{\theta ,\bar {\theta} \}_{\hat \star} = 1 {\hskip 2cm}\theta \hat \star \theta= \bar \theta \hat \star\bar {\theta}= 0
\end{equation}
reflecting the correspondence with anti-commutation relations of the fermionic operators $f^+$ and $f^-$. Note that one can  obtain also a star analogue of the $su(2)$ algebra using the noncommutative Gra\ss mann variables. Indeed, the functions
\begin{equation}
j_{+} = {\cal P}_{1,0}(\theta,\bar {\theta}) = \bar \theta{\hskip 1cm}
j_{-} = {\cal P}_{0,1}(\theta,\bar {\theta}) = \theta{\hskip 1cm}
j_{3} = {\cal P}_{1,1}(\theta,\bar {\theta}) = \bar {\theta}\theta
\end{equation}
satisfy the relations
\begin{equation}
[j_{-} , j_{+}]_{\hat \star}= 2 j_{3} {\hskip 2cm}[j_{\pm} , j_{3}]_{\hat \star}= \pm  j_{\pm}
\end{equation}
The Casimir operator is defined by
\begin{equation}
C = \frac{1}{2}(j_{-}\hat \star j_{+} + j_{+}\hat \star j_{-}) + j_{3}\hat \star j_{3}
\end{equation}
By a direct computation , one obtain $C = \frac{1}{4}$.\\
Now, we consider the bosonic oscillator described by the Hamiltonian
\begin{equation}
H_b = b^{+}b^{-}
\end{equation}
where $b^{+}$ and $b^{-}$ are the standard creation and annihilation operators. Using the coherent states of the harmonic oscillator
\begin{equation}
\vert z \rangle = e^{- \frac{1}{2}z\bar z} \sum_{n=0}^{\infty}\frac {z^{n}}{\sqrt{n!}} \vert n \rangle
\end{equation}
The star product of two functions ${\cal A}(z , \bar z) = \langle z \vert A \vert z \rangle$ and  ${\cal B}(z , \bar z)= \langle z \vert B \vert z \rangle$ ( $A$ and $B$ operators acting on ${\cal H}_b = \{ \vert n \rangle ; n \in {\cal N}\}$) is defined by [10]
\begin{equation}
\bar \star = e^{\overleftarrow{\frac{\partial}{\partial z}}\overrightarrow{\frac{\partial}{\partial \bar z}}}
\end{equation}
This star product is the so-called Voros Product. Every operator $A$ can be expanded in the occupation number operator basis $\{ \vert n \rangle \langle m \vert \}$ and for which we associate the functions
\begin{equation}
\Phi_{mn}(z , \bar z) = \langle z \vert  n \rangle \langle m \vert  z \rangle =
e^{-z \bar z} \frac {{\bar z}^{m}}{\sqrt{m!}} \frac {z^{n}}{\sqrt{n!}}
\end{equation}
satisfying the completion relation $\sum_{m=0}^{\infty} \Phi_{mn}(z , \bar z) = 1$ and the orthogonality property
\begin{equation}
\Phi_{m,n}(z , \bar z) \bar \star \Phi_{m',n'}(z , \bar z) = \delta_{m,n'} \Phi_{m ,n'}(z , \bar z)
\end{equation}
The $\Phi_{0,0}(z , \bar z)$ provides a  star vacuum
\begin{equation}
z \bar \star \Phi_{0,0}(z , \bar z) = \Phi_{0,0}(z , \bar z) \bar \star \bar z = 0
\end{equation}
We have also
\begin{equation}
z \bar \star \Phi_{n,m}(z , \bar z) = \sqrt{n}\Phi_{n-1,m}(z , \bar z)
\end{equation}
\begin{equation}
\bar {z} \bar \star \Phi_{n,m}(z , \bar z) = \sqrt{n+1}\Phi_{n+1,m}(z , \bar z)
\end{equation}
\begin{equation}
\Phi_{n,m}(z , \bar z) \bar \star z = \sqrt{m+1}\Phi_{n,m+1}(z , \bar z)
\end{equation}
\begin{equation}
\Phi_{n,m}(z , \bar z) \bar \star \bar {z} = \sqrt{m}\Phi_{n,m-1}(z , \bar z).
\end{equation}
The function ${\cal H}_{cla}$ associated to the Hamiltonian $H_b$ is
\begin{equation}
{\cal H}_{cla} = \langle z \vert H \vert z \rangle = z\bar z
\end{equation}
known ,in the coherent states language, as the identity action. One has also
\begin{equation}
{\cal H}_{cla} \bar \star \Phi_{n,m}(z , \bar z) = m \Phi_{n,m}(z , \bar z)
\end{equation}
\begin{equation}
\Phi_{n,m}(z , \bar z) \bar \star {\cal H}_{cla} = n \Phi_{n,m}(z , \bar z)
\end{equation}
At this stage, we have the main tools to introduce the star analogue of the superoscillaor. The supersymmetric oscillator is charactrized by the Hamiltonian
\begin{equation}
H_{susy} = b^{+}b^{-} + f^{+}f^{-}
\end{equation}
The supersymmetric character appears from the fact that the Hamiltonian $H_{susy}$ can be written as the anti-commutator $H_{susy} = \{Q^{-} , Q^{+}\}$ of the conserved supercharges $Q^{-}$ and $Q^{+}$
\begin{equation}
Q^{-} = b^{+}f^{-} {\hskip 2cm} Q^{+} = b^{-}f^{+}
\end{equation}
satisfying the nilpotency conditions $(Q^{-})^{2} = (Q^{+})^{2} = 0$ and commuting with the supersymmetric Hamiltonian. The operators $Q^+$, $Q^-$ and $H$ close the $Z_2$ graded algebra $sl(1/1)$. Since we are concerned with the superstar product in the supercoherent states framework, we consider the $Z_2$-graded Fock space
\begin{equation}
{\cal H} = {\cal H}_{b}\otimes {\cal H}_{f} = \{ \vert n , s \rangle ; n \in N ,  s = 0 , 1\}
\end{equation}
Using the group theoretical approach, based on the Weyl-Heisenberg supergroup $W$, the action of a unitary representation of $W$ leads to the supercoherent states [16-17]
\begin{equation}
\vert z , \theta \rangle = ( 1 - \frac{1}{2}\bar {\theta}\theta )( \vert z , 0 \rangle + \theta \vert z , 1 \rangle ) = \vert z \rangle \otimes \vert \theta \rangle
\end{equation}
where the bosonic and fermionic variables are commuting. Following the method, presented in the second section leading to the Gra\ss mann star product, the superstar product is defined by means of the supercoherent states (98) as
\begin{equation}
\star_{susy} = e^{\overleftarrow{(\frac{\partial}{\partial z}} \overrightarrow{\frac{\partial}{\partial \bar z}} + \overleftarrow{\frac{\partial}{\partial \theta}} \overrightarrow{\frac{\partial}{\partial \bar {\theta}}})}
\end{equation}
The functions associated to $H_{susy}$ and supercharges $Q^{+}$, $Q^{-}$ are given by
\begin{equation}
h_{susy} = \langle z , \theta  \vert H_{susy} \vert z , \theta \rangle  = \bar {z}z + \bar {\theta}\theta
\end{equation}
\begin{equation}
q^{-} = \langle z , \theta  \vert Q^{-} \vert z , \theta \rangle  = z \bar {\theta}
\end{equation}
\begin{equation}
q^{+} = \langle z , \theta  \vert Q^{+} \vert z , \theta \rangle  = \bar{z} \theta
\end{equation}
Moreover, It is easy to verify the following relations
\begin{equation}
q^{-}{\star_{susy}}q^{+} + q^{+}{\star_{susy}}q^{-}  = h_{susy} {\hskip 2cm} q^{\pm}\star_{susy} h_{susy} =  h_{susy}\star_{susy} q^{\pm}
\end{equation}
and
\begin{equation}
q^{+}\star_{susy}q^{+} = q^{-}\star_{susy}q^{-} = 0
\end{equation}
which are similar to ones defined the $sl(1/1)$ superalgebra where the ordinary (anti)-commuations are replaced by $\star$ ones.
\section{\bf Concluding remarks}
To close this note, recall that the works [10-13] related to star products and Moyal brackets, formulated in the coherent states framework, are exclusively of bosonic type. In the best of our knowledge, the Gra\ss mannian case involving fermionic coherent states has not been considered elsewhere. This is most probably due to the deficiency of a widely accepted convincing description of classical systems by means of Gra\ss mann algebras. An additional reason may be arise from the unconventional techniques for integration and differentiation of gra\ss mann variables. In this work, we tried to fill this gap and purposed a non anti-commutative Gra\ss mann star product. The anti-commutative limit $\hbar \to 0$ (classical limit in some sense) is easily recovered by a simple rescaling of the variables $\theta \to \frac{1}{\sqrt{\hbar}}\theta$. It has been shown that each unitary transformation admits a unique representation in the noncommuative space spanned by Gra\ss mann variables and the covariance holds only and only if the transformation is canonical. The supersymmetric extension of the Voros star product is also considered  thanks to the supercoherent states. Finally, we believe that the results of this work can be used to define star or super star products linked to Lie and super Lie algebras, as for instance $su(p,q)$ and $Osp(p/q)$, and we hope to report on this subject in a future paper.
{\vskip 1.0cm}
{\bf Acknowledgements}:
The author would like to thank the Condensed matter Section of the Abdus Salam-ICTP for hospitality.\\
\vfill\eject

\end{document}